\documentclass{PoS}

\pdfoutput=1

\usepackage{epsfig}
\usepackage{graphicx}
\usepackage{amsfonts}
\usepackage{amsmath}
\usepackage{amssymb}
\usepackage{amsthm}
\usepackage{latexsym}
\usepackage{slashed}
\usepackage{array}
\usepackage{cite}

\newcommand{\be}{\begin{equation}}
\newcommand{\ee}{\end{equation}}
\newcommand{\ba}{\begin{eqnarray}}
\newcommand{\ea}{\end{eqnarray}}
\newcommand{\bi}{\begin{itemize}}
\newcommand{\ei}{\end{itemize}}
\newcommand{\tr}{{\rm Tr\,}}
\renewcommand{\vec}[1]{{\bf #1}}

\title{Determination of the Wilson ChPT low energy constant $c_2$}

\ShortTitle{Determination of $c_2$}

\author{\speaker{Fabio Bernardoni}\thanks{We thank O. B\"ar, M. L\"uscher, S. Necco and F. Virotta for useful discussions.}\\
        NIC, DESY, Zeuthen\\
        E-mail: \email{fabiob@ifh.de}}

\author{John Bulava\\
        NIC, DESY, Zeuthen\\
        E-mail: \email{john.bulava@desy.de}}
        
\author{Rainer Sommer\\
        NIC, DESY, Zeuthen\\
        E-mail: \email{sommer@ifh.de}}

\abstract{Following a suggestion by Aoki and B\"ar, $c_2$ can be extracted by 
analyzing volume effects in 2 pion states. To this end we consider 
renormalized ratios of four point to two point correlation functions. We 
present the results from various CLS lattices, with pion masses ranging from 
280 to 450 MeV and lattice spacings of 0.07 fm and 0.08 fm. This low energy 
constant is useful to understand discretization effects in chiral 
extrapolations with Wilson fermions, especially for quantities which vanish in the chiral limit, like the 
pion mass. Since our procedure is computationally cheap and straightforward, it is recommended as a routine study for any Wilson-type simulation, as a check of discretization effects.
\vspace{2cm}
\flushright{DESY 11-209\\SFB/CPP-11-67}

}

\FullConference{ The XXIX International Symposium on Lattice Field Theory - Lattice 2011\\
July 10-16, 2011\\
Squaw Valley, Lake Tahoe, California}

\begin{document}

\section{Introduction}
For a few years we have had algorithms that allow simulation at small quark masses $m$ in large volumes, such that $M_\pi L \gtrsim 4$. As a consequence of 
this progress, and since volume effects are exponentially suppressed in $M_\pi L$, they are presently kept at the level of $1 \%$. On the other hand, discretization effects obey power law scaling, which in the case of $O(a)$ improved 
Wilson fermions amount to $O(a^2)$ effects. Most importantly, for all kinds of 
Wilson fermions, the dominant discretization effects are related to the 
breaking of chiral symmetry. As a consequence, they become more relevant as we 
try to reach lower quark masses at fixed $a$. It is then clear that the 
understanding and removal of discretization effects is essential.

Recent studies~\cite{Schaefer:2010hu} however have shown a severe critical 
slowing down which makes it unrealistic to simulate lattices with 
$a\lesssim 0.05$ fm. While the critical exponent can be reduced from $\sim5$ 
to 2 by using open boundary conditions~\cite{Luscher:2011kk}, to reduce the 
lattice spacing remains challenging and an understanding of the discretization 
effects is necessary to establish trustworthy results.

One framework to study the discretization effects is provided by Wilson ChPT 
(WChPT)~\cite{Sharpe:1998xm}, the chiral effective theory of Wilson LQCD. In 
this effective theory the cutoff effects are parametrized at a given order in the chiral expansion by a finite number of Low Energy Constants (LECs), among 
which we find $c_2$.
In terms of the physical and unphysical LECs, Wilson ChPT predicts the 
scaling of pion observables with respect to $m$ and $a$. In this context, 
for one class of observables, $c_2$ plays an important role as it appears in 
the LO lagrangian.

In this talk we present our results and strategy to put at least an upper 
limit on $|c_2|$ and eventually establish its sign. We also discuss the 
relevance that the values found would have on the $m$--dependence of some physical observables.

\section{Wilson ChPT}

%The procedure to build WChPT is sketched in Table~ \ref{sketch}. 
The first step to build WChPT is to collect all the operators in the continuum
Symanzik effective theory that are compatible with the symmetries of Wilson 
LQCD 
up to a certain dimension (i.e. power in $a$). These then have to be matched to
the chiral effective theory; they must be rewritten in terms of pion fields. 
Once a power counting for $a$ relative to $m$ and $p$ has been established, 
the spurion technique which was already used to introduce chiral symmetry 
breaking effects due to $m$ can be extended to treat cutoff effects. 

Our aim is to demonstrate that cutoff effects are not strong enough to distort 
the 
physical chiral scaling of the observables in which we are interested. As a 
consequence, we measure on our coarser lattices with $a=0.075\mathrm{fm}$  and 
$a=0.065\mathrm{fm}$ and assume a conservative power counting for $a$ (the so called LCE regime) which assumes:
\be
m \sim a^2\Lambda_{QCD}^3 \quad a\simeq 0.07 \mbox{ fm} \Rightarrow a^2\Lambda_{QCD}^3 \sim 10 \mbox{ MeV} \,.
\label{powcou}
\ee
This counting is expected to be the proper one for quark masses around $10\mathrm{MeV}$. However large or small coefficients in the expansion could shift this 
limit up or down so a practical test is necessary. We also assume (which is 
true in our case) that $O(a)$ improvement has been carried out 
nonperturbatively. 

With the power counting in eq.~\eqref{powcou}, the LO chiral Lagrangian is:
\be
{\cal L}_\chi = \frac{F^2}{4}\tr[\partial_\mu U \partial_\mu U^\dagger]-\frac{\Sigma m}{2}\tr [U+U^\dagger]+  \frac{c_2 F^2a^2}{16}\tr[U+U^\dagger]^2 
\label{lag}
\ee
where $c_2$ is a new unphysical LEC. One can show that the pion mass is corrected at LO:
\be
M_\pi^2= M_0^2+ O(M_0^4)  \qquad M_0^2 = \frac{2\Sigma m}{F^2} - 2 c_2a^2  
\label{M0}
\ee
and that a new phase pattern emerges \cite{Aoki:1983qi}, whose properties are controlled by 
the 
sign of $c_2$ \cite{Sharpe:1998xm}.  For Wilson fermions, if $c_2>0$ it is 
possible to reach a phase where flavor and parity are spontaneously broken 
by decreasing 
the quark mass at fixed $a$. Since the transition is of second order, the 
pions are massless on the boundary. Conversely, if $c_2$ is negative, the 
transition is first order, flavor and parity remain unbroken, and there are 
no massless pions at finite $a$.

We now possess  a tool to predict the scaling on $m$ and $a$ (up to a given order) in terms of the physical LECs of ChPT plus some new ones (including $c_2$). 
It should be noted that $m$ appearing in the Lagrangian (Eq.~\eqref{lag}) is just a bare parameter, which has no immediate relation to a well-defined quark (or pion) mass of the lattice theory. One example of such a quark mass is the 
PCAC mass, $m_{PCAC}$. Before the WChPT formulas can be used, it is necessary 
to find a relation between $m$ and $m_{PCAC}$  (or $M_\pi$) that allows to rewrite the observable of interest ${\cal O}$:
\be
\tilde{\cal O}(m_{PCAC}) = {\cal O}(m(m_{PCAC}))\quad \mbox{or} \quad \tilde{\cal O}(M_\pi) = {\cal O}(m(M_\pi) \,.
\ee
For example, one can obtain the pion mass as a function of $M_0$ (or $m$ through Eq.~\eqref{M0}) \cite{Aoki:2008gy}\footnote{Here and in the following we do not write the analytic terms (polynomial in $M_\pi^2$ ) explicitly for simplicity, but they have also been calculated.}:
\be
M_{\pi}^2=M_0^2 \left[1+\frac{M_0^2+10c_2a^2}{64\pi^2F^2}\ln \left(\frac{M_0^2}{\tilde{\mu}_1^2}    \right) +O(a^2,m)\right]+O(a^4) \,,
\ee
but this formula cannot be used for extrapolations until a relation between 
$M_0$ and a measurable mass is found. Unfortunately we know  just the LO relation between $m$ and $m_{PCAC}$~\cite{Bar:2008th}:
\be
m_{PCAC}= \left(m-\frac{c_2 a^2F^2}{\Sigma}   \right) (1+O(a^2,m)) \,.
\label{mPCAC}
\ee
After the adequate substitutions are performed, one finds that observables 
can be separated in two classes, according to how much they are affected 
by discretization effects.
\begin{itemize}
\item Observables that are nonzero in the chiral limit, such as the pion decay constant. Cutoff effects appear here just as analytic contributions and at NLO, both in terms of $M_\pi$ and $m_{PCAC}$:
\be
F_\pi^2 = F^2 \left[1-\frac{M_{\pi}^2}{8\pi^2F^2} \ln \frac{M_{\pi}^2}{\tilde{\mu}_4^2}  +O(a^2,m) \right]
\ee
\item Observables that vanish in the chiral limit, such as the pion mass or scattering lengths. Cutoff effects appear at LO, they change the coefficient of the chiral logarithms in such a way that the expansion breaks down at low $M_\pi$. Here we show the S-wave scattering length for $I=2$, $a_0^2$, as an example, that will also be useful in the following:
{\small 
\be
M_{\pi}a_0^2 =-\frac{M_{\pi}^2}{16\pi
 F_\pi ^2}\left[1+\frac{3M_{\pi}^2+12c_2a^2}{32\pi^2 F_\pi ^2}\ln \frac{M_{\pi}^2}{\tilde{\mu}_2^2}+O(a^2,m)\right]
 -\frac{2c_2a^2}{16\pi
 F_\pi ^2}\left[1+\frac{11c_2a^2-2M_\pi^2}{16\pi^2 F_\pi ^2}\ln \frac{M_{\pi}^2}{\tilde{\mu}_3^2}+O(a^2,m)\right] \,.
 \label{scatlen}
\ee
}
\end{itemize}

\section{Determination of $c_2$}

From Eqs.~\eqref{mPCAC} and \eqref{M0} it is apparent that in the relation $M_\pi^2(m_{PCAC})$, $c_2$ appears only at NLO. Conversely $c_2$ enters at LO 
in the relation between the S-wave scattering length and $M_\pi$, see 
Eq.~\eqref{scatlen}. The energy eigenvalues of interacting particles are 
affected in finite volume by a term proportional to the scattering length according to the L\"uscher formula:
\be
E_{2\pi}(\vec{p}=0)= 2 M_\pi +\Delta E  \qquad  \Delta E\equiv-\frac{4\pi a_0^I}{M_\pi L^3}+O(L^{-4})   
\ee
which opens up the possibility to measure $c_2$ through volume effects.

\begin{figure}
\begin{center}
 \includegraphics[width=8cm]{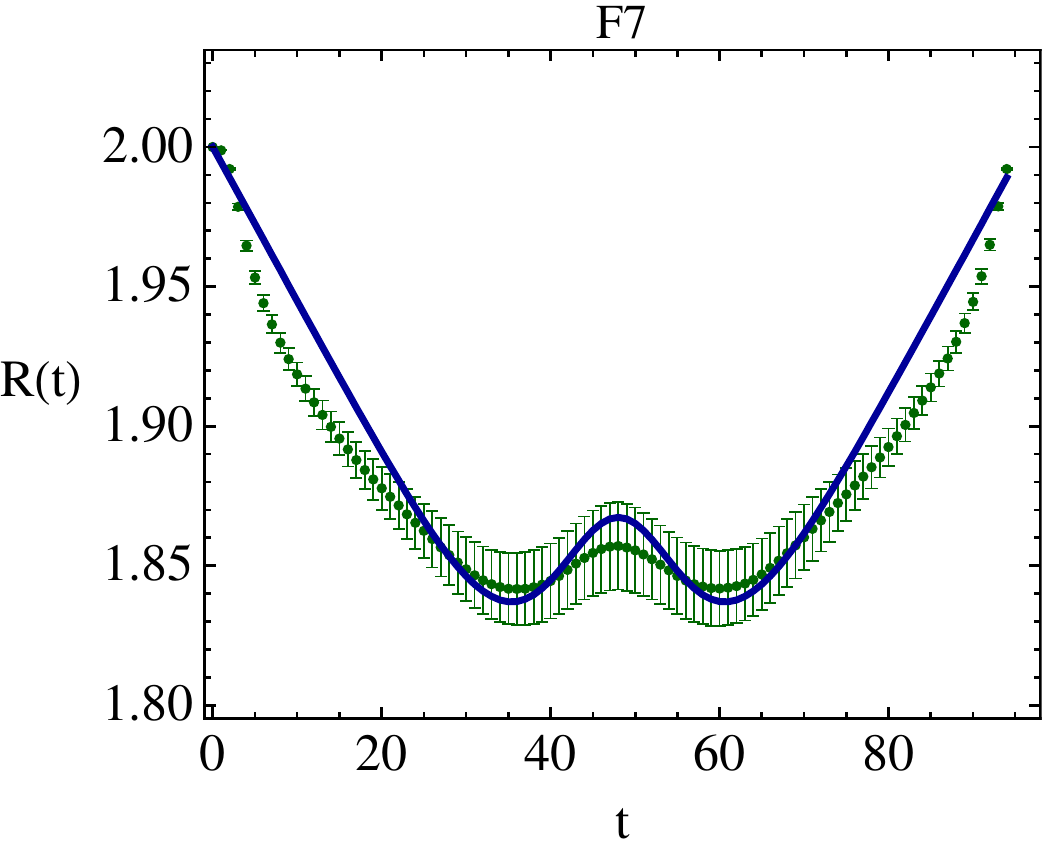}
 \end{center}
 \caption{R(t) for the F7 ensemble. 
 The blue curve represents the WChPT prediction.}
 \label{ratio}
\end{figure}

We consider the ratio of correlators:
\be
R(t) \equiv \frac{\langle \pi^+ (t)\pi^+ (t)  \pi^-(0) \pi^- (0 ) \rangle}{\langle  \pi^+ (t)  \pi^- (0)  \rangle^2}\xrightarrow[T\to \infty]{t\gg1/M} e^{-\Delta E t}
\ee
where $\pi^\pm (t) \equiv \sum_{\vec{x}}\pi^\pm (t,\vec{x})$. Apart from the 
QCD coupling and quark masses, $R(t)$ needs no renormalization
and at LO in ChPT it depends only on $F_\pi$, $M_\pi$ and $c_2$. The result 
for $R(t)$ on one of our lattices is shown in Fig.~\ref{ratio}.

\begin{figure}
\begin{center}
\includegraphics[width=6cm]{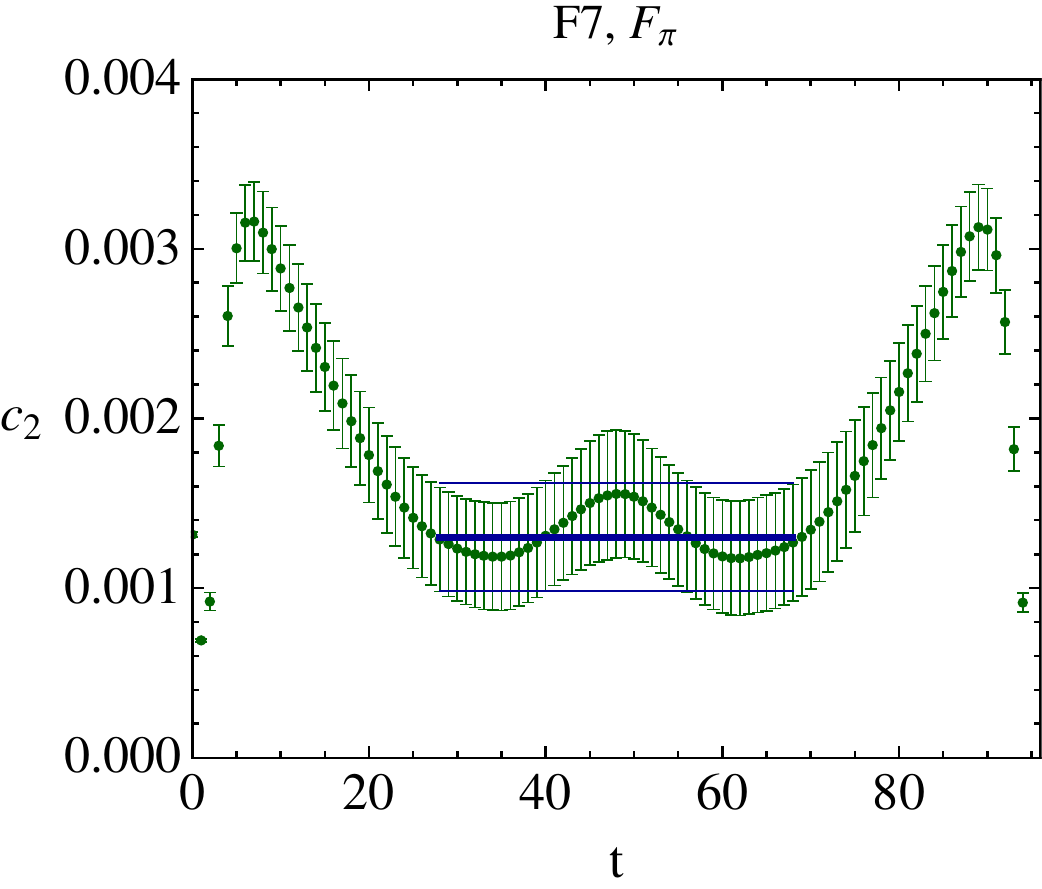}
\includegraphics[width=6cm]{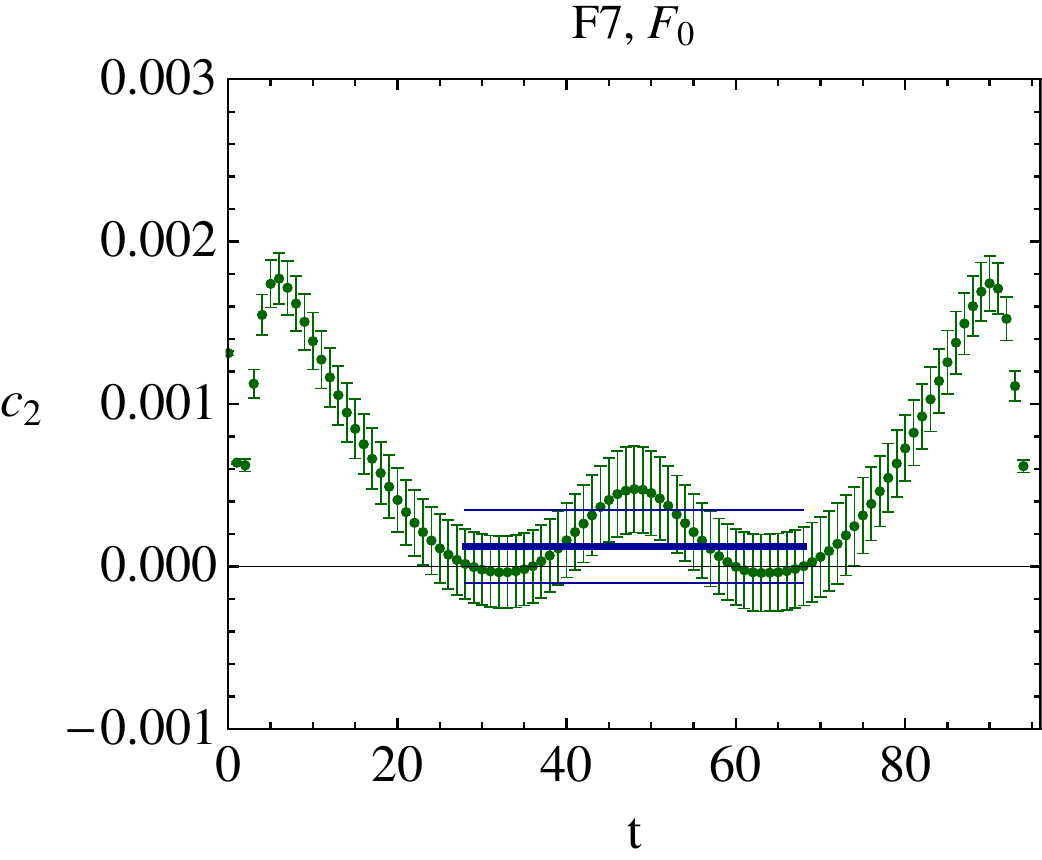}
\end{center}
\caption{The effective $c_2(t)$ with the observed decay constant $F_\pi$ (left) and its value in the chiral limit (right).}
\label{c2t}
\end{figure}

We have evaluated $R(t)$ on the coarser CLS lattices listed in table 
\ref{lattices} and in WChPT, in the LCE regime and finite volume. We evaluate 
all-to-all quark propagators stochastically, using 4 fully time diluted noise 
sources per configuration. Using $F_\pi$ and $M_\pi$ determined from 
the two point correlation functions, we invert the ChPT formula to obtain 
an effective $c_2(t)$. As shown in Fig.~\ref{c2t} we then average over a time 
interval $[t_{min},t_{max}]$ such that $t_{min} > k/M_\pi$, $(T-t_{max})> k/M_\pi$ for fixed $k={\rm O}(1)$. In practice we have checked that the result does not change significantly for $k \in [2,3]$ and in the following we report the results obtained for $k=2.5$. This definition converges to the LEC $c_2$ in the chiral limit. We treat errors coming from autocorrelations in a conservative way, using the techniques described in \cite{Schaefer:2010hu}. To estimate the 
systematic effect coming from truncating the chiral expansion at LO we compare 
the results using in the chiral formulas both 
the observed pion decay constant $F_\pi$ and its value in the chiral limit 
$F_0$, 
see Fig.~\ref{c2t}.

\begin{figure}
\begin{center}
\includegraphics[width=6cm]{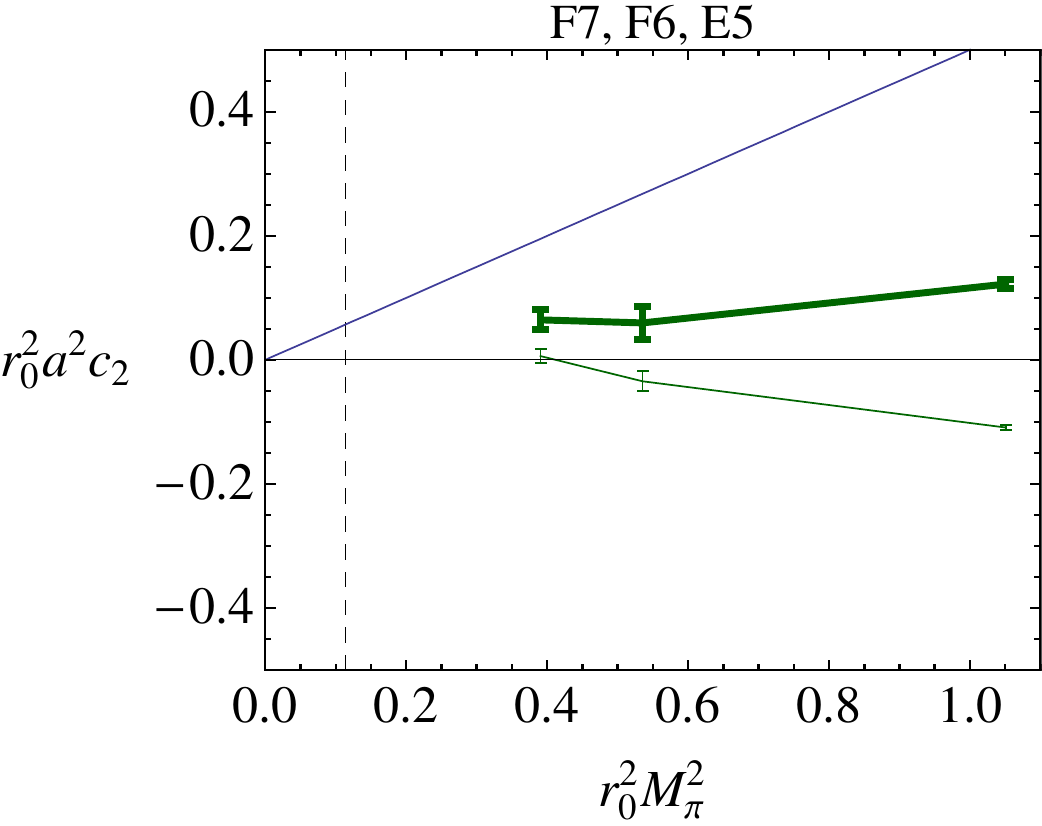}\includegraphics[width=6cm]{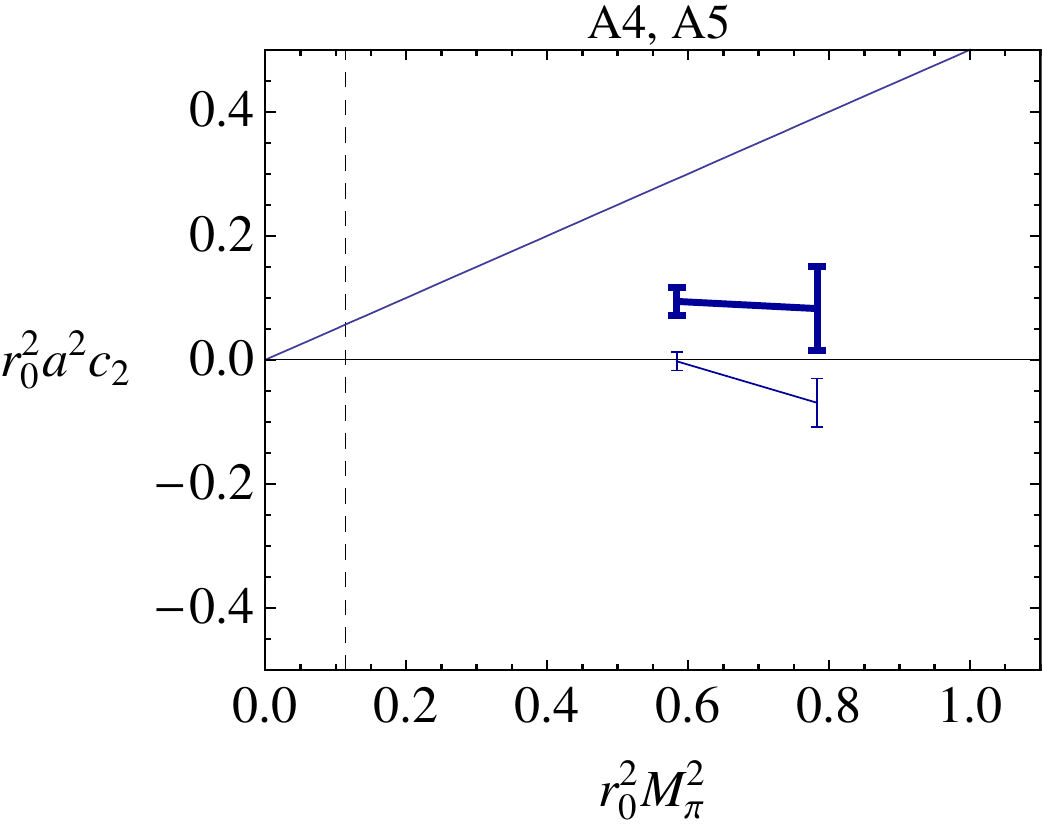}
\end{center}
\caption{Results for $c_2$ on lattices with $a=0.065$ fm (left) and $a=0.075$ fm (right). Thick lines corresponds to $F_\pi$ while thin lines corresponds to $F_0$. The dashed line indicates the physical point, the blue line corresponds 
to $2c_2a^2=M_\pi^2$, which is where the physical contribution to the 
scattering length is of the same magnitude of the one coming from 
cutoff effects.}
\label{summary}
\end{figure}

The results that we obtained are summarized in Fig.~\ref{summary}. Both determinations (using $F_\pi$ and using $F_0$) seem to converge to a positive value, 
and all our points lie far from the dangerous region (the blue line) 
where the physical (due to $M_\pi\ne0$) and unphysical (due to $a\ne0$) 
contributions to the S-wave $I=2$ scattering length are equal. It is 
interesting to mention that the ETMC collaboration found a negative 
value~\cite{Michael:2007vn}, which is not in contradiction with our 
result, since their action was different.

\begin{table}
\begin{center}
\begin{tabular}{|c|c|c|c|c|}
%\multicolumn{5}{|c|}{$\beta= 5.3$,  $V=48\times 24^3$, $a=0.0784(10)$ fm} \\ [0.1cm]
\hline
& & &  &\\
label    & $a   (fm)     $              &  $M_\pi$   (MeV) &  $M_\pi$L    &  $N_{\rm cfg}$  \\
\hline
& & & &  \\
A$_4$ & 0.075 & 380  &  4.1   &  100       \\ 	  
A$_5$ & 0.075 & 330  &  4.7   &  325       \\	
E$_5$ & 0.065 & 440  &	  4.7   &  999     \\   
F$_6$ & 0.065 & 310  &	  5      &  300     \\  
F$_7$ & 0.065 & 270  &   4.2  &  600       \\                          
\hline
\end{tabular}
\end{center}
\caption{Lattices on which $R(t)$ was computed. Lattices used for physical 
measurements have lattice spacings as small as $a=0.05$ fm. The lattice 
spacing was set from $f_K=155$ MeV, see contribution by M. Marinkovic at this conference.}
\label{lattices}
\end{table}

\section{Conclusions and Outlook}

Our aim is to find whether or not we expect large cutoff effects in our 
simulations. Considering that we have performed tests on our coarser lattices 
and that we have assumed a pessimistic power counting, Fig.~\ref{summary} is a 
good indication that cutoff effects are small in our observables and the 
physical scaling with $M_\pi$ (or $m_{PCAC}$) is not significantly 
distorted for the masses that we are considering. Our procedure is cheap and 
simple, and will be repeated in the future for all ensembles. 

We have significant errors, coming mainly from the chiral truncation, that 
make it difficult to establish a conclusive value for $c_2$, although our 
data at $a=0.065$~fm point to a rather small and possibly positive value.

In the near future we will apply the same procedure to lattices with smaller 
masses than the ones so far considered. These additional points may help to 
establish the sign of $c_2$. If indeed $c_2\geq 0$ with our action, this 
would suggest adding a clover term to the action
used by the ETM collaboration also yields $c_2\geq0$ \cite{Boucaud:2008xu}. This is important, 
because in that context $M_{\pi^\pm}-M_{\pi^0}\propto c_2$ and a negative 
$c_2$ means large volume effects through $M_{\pi^0}<M_{\pi^\pm}$.

\bibliography{lattice}

\begin{thebibliography}{1}

\bibitem{Schaefer:2010hu}
ALPHA Collaboration, S. Schaefer, R. Sommer and F. Virotta,
\newblock Nucl.Phys. B845 (2011) 93, 1009.5228.

\bibitem{Luscher:2011kk}
M. Luscher and S. Schaefer,
\newblock JHEP 1107 (2011) 036, 1105.4749,
\newblock * Temporary entry *.

\bibitem{Sharpe:1998xm}
S.R. Sharpe and J. Singleton, Robert~L.,
\newblock Phys.Rev. D58 (1998) 074501, hep-lat/9804028.

\bibitem{Aoki:1983qi}
S. Aoki,
\newblock Phys.Rev. D30 (1984) 2653.

\bibitem{Aoki:2008gy}
S. Aoki, O. Bar and B. Biedermann,
\newblock Phys.Rev. D78 (2008) 114501, 0806.4863.

\bibitem{Bar:2008th}
O. Bar, S. Necco and S. Schaefer,
\newblock (2008), 0812.2403,
\newblock %%CITATION = 0812.2403;%%.

\bibitem{Michael:2007vn}
ETM Collaboration, C. Michael and C. Urbach,
\newblock PoS LAT2007 (2007) 122, 0709.4564.

\bibitem{Boucaud:2008xu}
ETM collaboration, P. Boucaud et~al.,
\newblock Comput.Phys.Commun. 179 (2008) 695, 0803.0224.

\end{thebibliography}
\bibliographystyle{h-elsevier}
\end{document}